\documentclass[nofootinbib,aps,12pt]{revtex4-1}
\usepackage{graphicx}
\usepackage{cancel}
\usepackage{textcomp}
\usepackage{amsmath}
\usepackage{bm}
\usepackage{epsfig}
\usepackage{color}
\usepackage{epstopdf}
\usepackage{dcolumn}
\def\beq{\begin{equation}}
\def\eeq{\end{equation}}

\def\be{\begin{equation}}
\def\ee{\end{equation}}
\def\bea{\begin{eqnarray}}
\def\eea{\end{eqnarray}}

\begin{document}
\title{\Large Searching for NLSP Sbottom at the LHC}
\bigskip
\author{M. Adeel Ajaib\footnote{email: adeel@udel.edu}}
\author{Tong Li\footnote{email: tli@udel.edu, communication author}}
\author{Qaisar Shafi\footnote{email: shafi@bartol.udel.edu}}
\address{
Bartol Research Institute, Department of Physics and Astronomy,
University of Delaware, Newark, DE 19716, USA}
\date{\today}

\begin{abstract}
We study the collider phenomenology of sbottom-bino co-annihilation scenario at both the 7 TeV and 14 TeV LHC. This co-annihilation scenario requires that the NLSP sbottom and LSP bino masses are apart by no more than about 20\% or so, and for $M_{\tilde{b}_1}>M_b+M_{\tilde{\chi}_1^0}$, the sbottom decays exclusively into $b+\tilde{\chi}_1^0$. We propose a search for sbottom pairs through $b\bar{b}$ plus missing energy. By scanning the mass parameters $M_{\tilde{b}_1}$ and $M_{\tilde{\chi}_1^0}$, we investigate the discovery limits of sbottom and bino in the $M_{\tilde{b}_1}-M_{\tilde{\chi}_1^0}$ plane with at least $5\sigma$ significance at the LHC, for varying integrated luminosities. It is shown that with at least 5 fb$^{-1}$ luminosity, the 7 TeV LHC can explore a narrow region satisfying the $20\%$ co-annihilation condition. For the 14 TeV LHC with 10 (100) fb$^{-1}$ luminosity, the discovery limit of $M_{\tilde{b}_1}$ is 360 (570) GeV. 
\end{abstract}
 \maketitle

\section{Introduction}
Low scale supersymmetry, augmented by an unbroken $Z_2$ matter (R-) parity, largely overcomes the gauge hierarchy problem encountered in the Standard Model (SM) and also provides a compelling cold dark matter candidate. In the mSUGRA/constrained minimal supersymmetric model (CMSSM)~\cite{Arnowitt:2006bb} , as well as in many other realistic models, the lightest neutralino (LSP) is stable~\cite{lsp} with a relic density that is compatible with the WMAP dark matter measurements~\cite{wmap}. To be compatible with the latter, and assuming thermal relic abundance, a pure Higgsino or wino LSP mass should be around a TeV or so~\cite{bino}, which is about an order of magnitude larger than the most sensitive discovery region ($\sim 100$ GeV) of the ongoing direct detection experiments. On the other hand, the small annihilation cross section of a pure bino LSP with mass of around 100 GeV does not permit one to easily reproduce the required relic dark matter abundance~\cite{bino}.


A variety of scenarios that enhance the bino annihilation cross section have been proposed. For instance, the LSP could be a suitable bino-Higgsino admixture with mass $\sim 100$ GeV and compatible WMAP relic density~\cite{Gogo} (and references therein). A somewhat different and well studied option is co-annihilation, which is realized in the CMSSM as bino-stau or bino-stop co-annihilation~\cite{coann1,coann2}. In this case the bino and the relevant NLSP (where NLSP stands for next to lightest supersymmetric particle) sfermion are sufficiently close together in mass, such that the ensuing co-annihilation processes in the early universe allow one to reproduce the desired bino relic density. Another interesting scenario of this kind is bino-gluino co-annihilation. This scenario is not possible in the CMSSM, but it has been implemented in models with non-universal gaugino masses~\cite{coann3}, and a class of (third family) Yukawa unified models~\cite{422}. The collider signatures of these three (stop, stau, gluino) co-annihilation scenarios have been discussed in the literature~\cite{stau,stop,gluino} and references therein.

A somewhat less well studied region of the MSSM parameter space is the so-called sbottom co-annihilation (or NLSP sbottom) scenario. This was proposed in the context of supersymmetric $SU(5)$ with $b-\tau$ unification and non-universal supersymmetry breaking scalar $\mathrm{mass}^2$ terms for the $\bf \bar{5}$ and $\bf 10$ matter multiplets~\cite{Profumo}. More recently, the NLSP sbottom scenario was re-considered in Ref.~\cite{ilc}, motivated by studies related to the International Linear Collider (ILC). Yet another suggestion, put forward in Ref.~\cite{Pallis}, is to employ non-universal gaugino masses. Be that as it may, in our study we do not specify any particular theoretical models to explore the sbottom-bino co-annihilation scenario.

A satisfactory implementation of the NLSP sbottom scenario with the correct WMAP compatible LSP relic density requires a relatively small mass difference, less than or of order 20\%, between the sbottom and the bino~\cite{Profumo,Pallis,ilc}. The sbottom decay into $b$-jet, $\tilde{b}_1\to b\tilde{\chi}_1^0$, then becomes the unique channel if the mass difference is larger than $b$ quark mass and suggests a search for NLSP sbottom via
\begin{eqnarray}
pp\to \tilde{b}_1\tilde{b}_1^\ast X\to b\bar{b}+\cancel{E}_T.
\end{eqnarray} 
This search was performed with the first Run II data in the D0 experiment, but at that time the $b$-tagging techniques were not available. Other groups have studied the detection possibility at the Tevatron~\cite{jose} and ILC~\cite{ilc} but, to the best of our knowledge, the analysis for LHC is still lacking. The di-$b$ jet events is also the leading search channel for the SM-like Higgs boson, but the identification criteria is different.

This paper is organized as follows. In section II we study, without 
relying on any specific NLSP sbottom model, the collider phenomenology 
of this scenario. We discuss here the signal properties and the relevant 
backgrounds. The NLSP sbottom discovery limits at the 7 and 14 TeV LHC 
are discussed in section III, and our conclusions are summarized in 
section IV.
 
\section{Sbottom Co-annihilation and NLSP Sbottom Search at the LHC}


With the requirement that
\begin{eqnarray}
{M_{\tilde{b}_1}-M_{\tilde{\chi}_1^0}\over M_{\tilde{\chi}_1^0}}\lesssim 20\%~,
\end{eqnarray}
the NLSP sbottom essentially decays into a bottom quark and LSP $\tilde{\chi}_1^0$, and consequently the relevant kinematic constraint is
\begin{eqnarray}
M_{\tilde{b}_1}>M_b+M_{\tilde{\chi}_1^0}
\end{eqnarray}
if this two-body decay is allowed. Together with the requirement of sbottom-bino co-annihilation, we have the following $M_{\tilde{\chi}_1^0}$ region for a given sbottom mass:
\begin{eqnarray}
M_{\tilde{b}_1}/1.2<M_{\tilde{\chi}_1^0}<M_{\tilde{b}_1}-M_b.
\label{bound}
\end{eqnarray} 
We discuss below the discovery potential of NLSP sbottom and LSP $\tilde{\chi}_1^0$ at the LHC with the mass boundary in Eq.~(\ref{bound}).



\begin{figure}[tb]
\begin{center}
\includegraphics[scale=1,width=8cm]{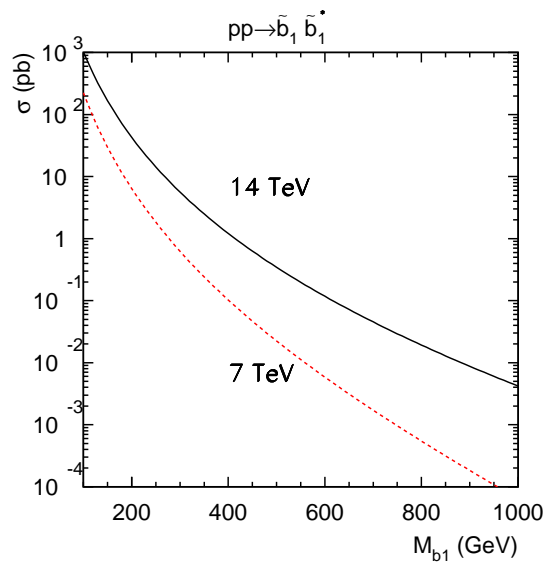}
\end{center}
\caption{The total cross section of sbottom pair production at the 7 TeV (dashed) and 14 TeV (solid) LHC. The renormalization and factorization scales are taken to be the sbottom mass.} 
\label{sbottom}
\end{figure}

The current bound on the sbottom mass comes from LEP II, namely $M_{\tilde{b}_1}>90$ GeV~\cite{LEP}. The pair production rate of sbottom with $\mathcal{O}(100)$ GeV mass is around the pico-barn level. For NLSP sbottom, we focus on the sbottom pair production followed by the unique two-body decay $\tilde{b}_1\to b\tilde{\chi}_1^0$. Namely, 
\begin{eqnarray}
pp\to \tilde{b}_1\tilde{b}_1^\ast X\to b\bar{b}+\cancel{E}_T,
\end{eqnarray}
induced by gluon fusion and $q\bar{q}$ annihilation processes. In principle, there also exist same sign sbottom pair productions $\tilde{b}_1\tilde{b}_1$ and $\tilde{b}_1^\ast \tilde{b}_1^\ast$. They arise from third generation parton scattering with $t$-channel neutralino exchange. Due to the smallness of the parton distribution functions (PDFs) of third generation sea quarks at large $p p$ collision energy, we ignore this contribution in our study.
Therefore, the sbottom production and subsequent decay only depend on the two mass parameters $M_{\tilde{b}_1}$ and $M_{\tilde{\chi}_1^0}$. In Fig.~\ref{sbottom} we show the total cross section of sbottom pair production at the 7 TeV (dashed) and 14 TeV (solid) LHC.

The decays, on average, of the relatively long lived $B$-mesons in the final states take place $\mathcal{O}$(mm) away from the primary interaction vertex. With the detection at the so-called secondary vertex, tagging jets with decaying $B$-mesons significantly reduces the QCD jet background. The $b$-jet production in the SM is mainly due to gluon splitting, and so the $b$-jets always arise as pairs. We therefore propose to tag two $b$-jets both in the signal and background events, and the SM $b\bar{b}$ production would become the leading irreducible background. 
Besides the $b$ quark, the dark matter particles $\tilde{\chi}_1^0$ also appear in sbottom decays. The missing transverse energy $\cancel{E}_T$ is another characteristic feature of the signal. The irreducible SM background for $\cancel{E}_T$ is from $Z$ production, with the branching fraction of $Z$ invisible decay $(Z\to \nu \bar{\nu})$ of 20\%. The other source of $b$-jet production in the SM is from top quark decay originating from top pair production. It can mimic our signal if both $W$'s decay leptonically and the charged leptons are not detected, which is assumed to occur if the leptons are too soft with transverse momentum $p_T^\ell<10$ GeV and are not in central range with pseudo-rapidity $|\eta_\ell|>2.5$. There also exist reducible backgrounds due to other jets being mis-identified as $b$-jets. The SM backgrounds we consider in our study are then
\begin{eqnarray}
b\bar{b} \ ; \ b\bar{b}Z,jjZ \  \ {\rm with} \ BR(Z\to \nu\bar{\nu})=20\% \ ; \ t\bar{t}\to b\bar{b}\ell^+\ell^-\nu\bar{\nu} \ \ {\rm with \ undetected \ leptons}.
\end{eqnarray}

We generate the SM background events with Madgraph/Madevent~\cite{Madgraph} and pass them into Pythia~\cite{Pythia} for parton shower and hadronization. Due to the uncertainty of mis-measurement in jet energy or momentum in the detector, the $b\bar{b}$ events without $Z$ can also induce missing energy. The Pythia output is then fed into the fast detector simulation PGS-4~\cite{PGS}, in order to simulate the important detector effects. The $b$-tagging efficiency and mis-tagging rate in PGS-4 are based on the Technical Design Reports of ATLAS and CMS, and we use the default values in our analysis. The following basic kinematical cuts on the transverse momentum ($p_T$), the pseudo-rapidity ($\eta$), and the separation in the
azimuthal angle-pseudo rapidity plane ($\Delta R=\sqrt{(\Delta
\phi)^2+(\Delta \eta)^2}$) between two jets have been employed for
jet selection~\cite{lhc}: \beq p^{j}_{T} > 15~{\rm GeV}, |\eta_{j}|<
2.0~, \Delta R_{jj}>0.4. \label{basic} \eeq

\begin{figure}[tb]
\begin{center}
\includegraphics[scale=1,width=8cm,height=5.8cm]{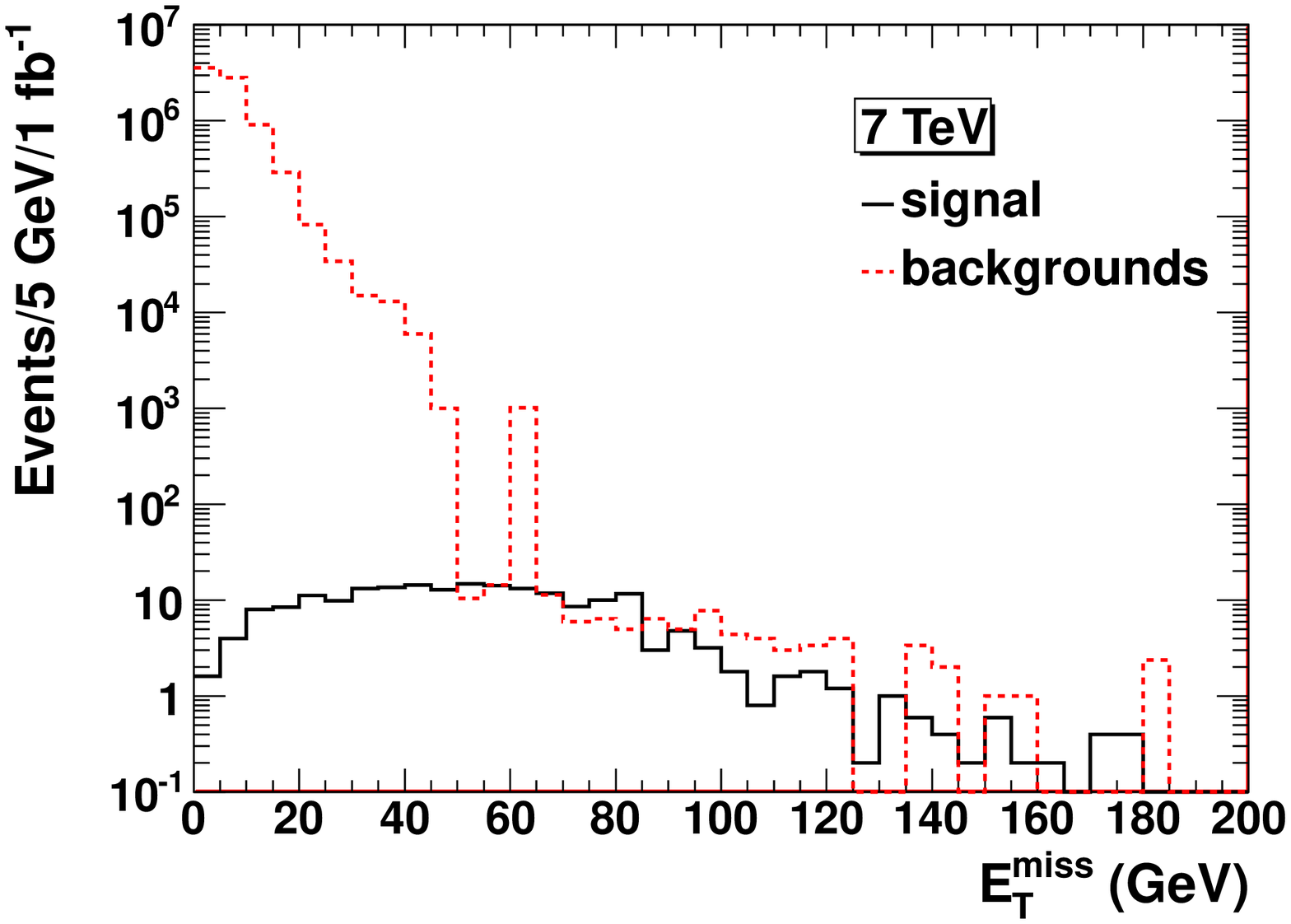}
\includegraphics[scale=1,width=8cm,height=5.8cm]{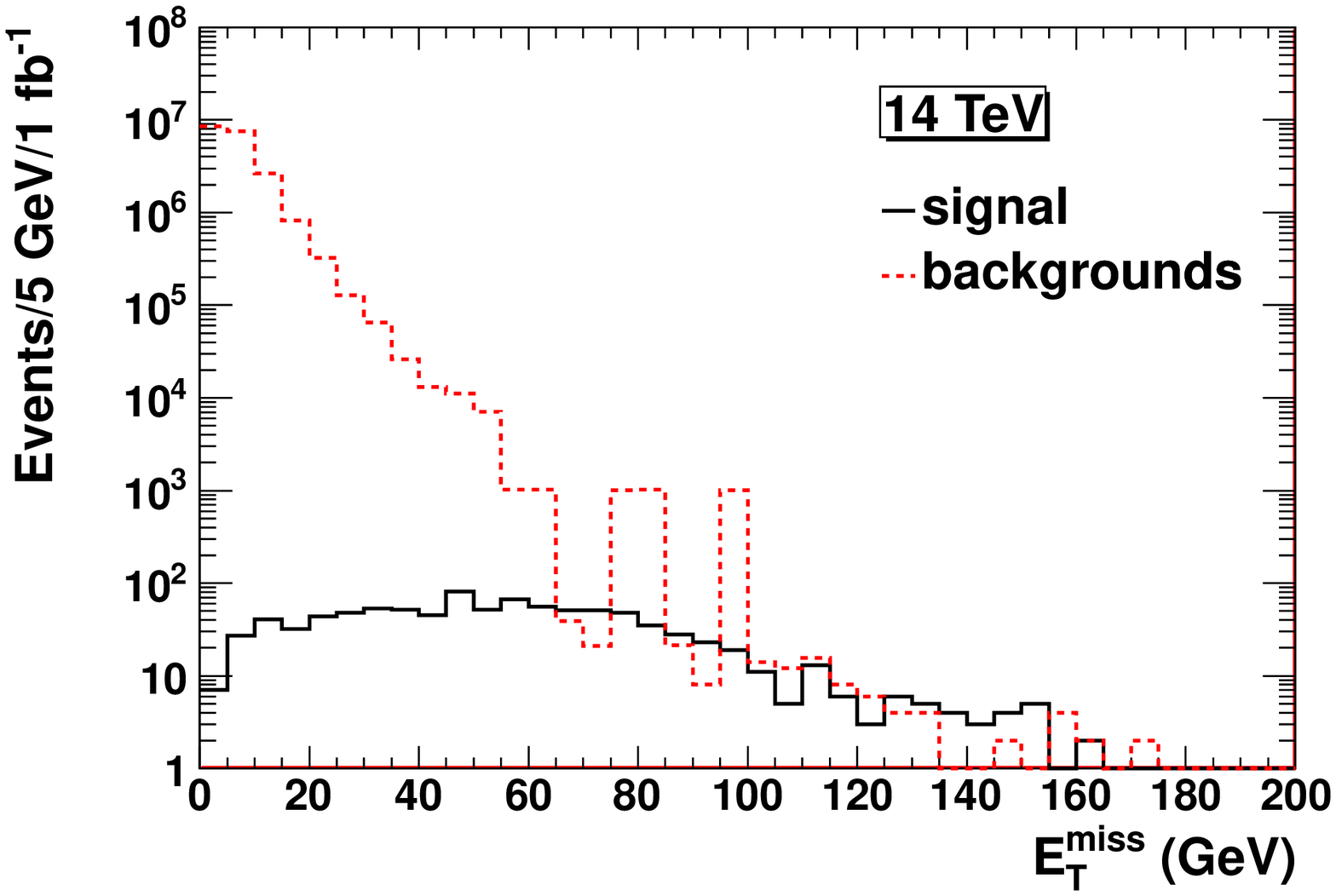}
\end{center}
\caption{$\tilde{b}_1\tilde{b}_1^\ast$ signal and SM background events vs. transverse missing energy $\cancel{E}_T$ at 7 TeV (left) and 14 TeV (right) LHC with 1 fb$^{-1}$ luminosity.  The masses of $\tilde{b}_1$ and $\tilde{\chi}_1^0$ are assumed to be 200 GeV and 150 GeV respectively.} \label{met}
\end{figure}

\begin{figure}[tb]
\begin{center}
\includegraphics[scale=1,width=8cm,height=5.8cm]{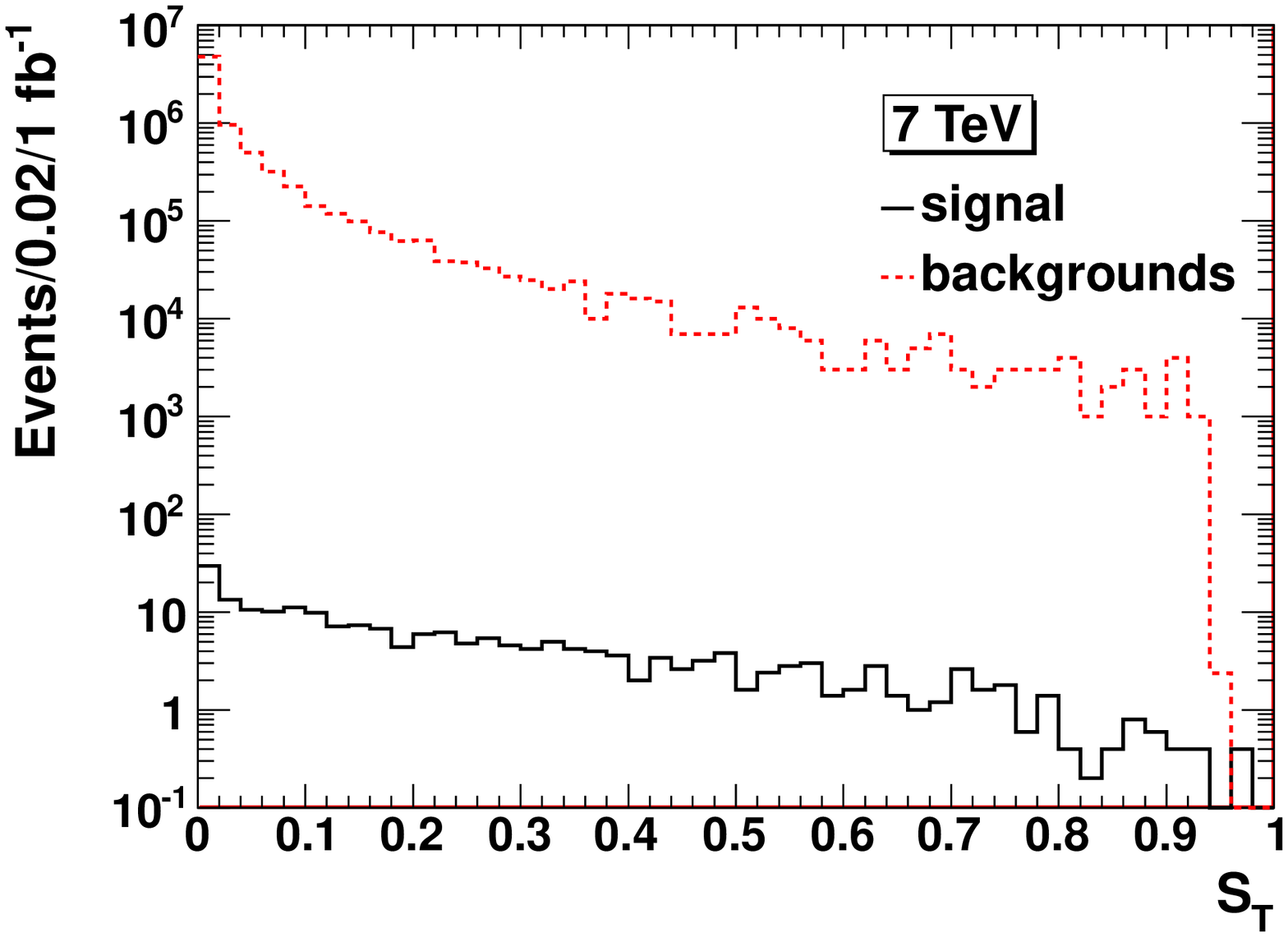}
\includegraphics[scale=1,width=8cm,height=5.8cm]{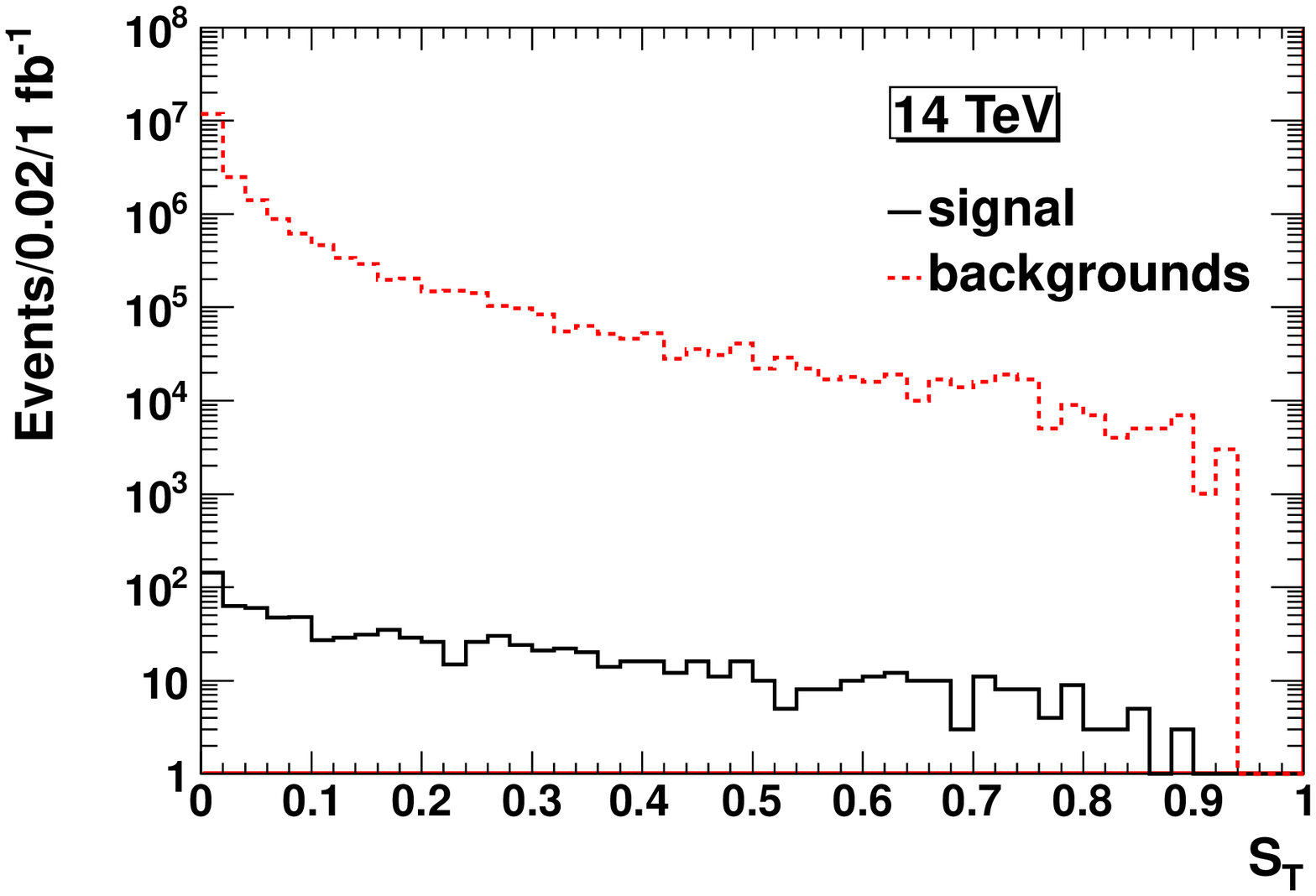}
\end{center}
\caption{$\tilde{b}_1\tilde{b}_1^\ast$ signal and SM background events vs. transverse sphericity $S_T$ at the 7 TeV (left) and 14 TeV (right) LHC with 1 fb$^{-1}$ luminosity.  The masses of $\tilde{b}_1$ and $\tilde{\chi}_1^0$ are assumed to be 200 GeV and 150 GeV respectively.} \label{st}
\end{figure}

For the SUSY signal we also use Pythia to generate events and then forward to PGS-4 to smear the observed particles as well. The hardness of the $b$-jet in final states depends on the mass difference $\Delta M=M_{\tilde{b}_1}-M_{\tilde{\chi}_1^0}$. The larger the $\Delta M$ we impose, the harder the $b$-jet is, as also the missing transverse energy $\cancel{E}_T$, in which case more events would pass the selection cuts. 
The $\cancel{E}_T$ is reconstructed according to the smeared $b$-jets. In Fig.~\ref{met} we show the $\cancel{E}_T$ distribution of the signal and backgrounds including basic cuts in Eq.~(\ref{basic}), and $b$-jet and mis-$b$ jet tagging efficiency from PGS. One can see that the $\cancel{E}_T$ of backgrounds is rather soft because the dominant background source is pure $b\bar{b}$, and its missing energy is just from small uncertainty of $b$-jets mis-measurement. We find that by requiring a significant $\cancel{E}_T$ cut, namely 
\begin{eqnarray}
\cancel{E}_T>40~{\rm GeV} \ {\rm for \ 7 \ TeV}, \ \ \cancel{E}_T>50~{\rm GeV} \ {\rm for \ 14 \ TeV},
\end{eqnarray}
the SM backgrounds can be significantly suppressed. However, $b\bar{b}$ is still the leading background and one order of magnitude larger than our signal because of its extremely large size to begin with. To further suppress the contribution from processes where the missing energy comes from jet energy mis-measurement, we require that $\cancel{E}_T$ is not parallel to any $b$-jets in events~\cite{jose}, namely 
\begin{eqnarray}
\Delta\phi(\cancel{E}_T,b)>0.2.
\end{eqnarray}
Furthermore, the $b\bar{b}$ events are expected to be back-to-back, and thus peaked at small transverse sphericity $S_T\sim 0$, as shown in Fig.~\ref{st}. Around $S_T\sim 0.2$, the behavior of backgrounds reaches a level two orders of magnitude lower than the peak value. We thus require~\cite{baer} 
\begin{eqnarray}
S_T>0.2
\end{eqnarray}
to reject $b\bar{b}$ events. After these selection cuts, the SM $b\bar{b}$ background can be completely eliminated. The remaining leading backgrounds are $b\bar{b}Z$ with invisible $Z$ decay and $t\bar{t}$ with undetected charged leptons. The efficiencies of different cuts for the signal $\tilde{b}_1\tilde{b}_1^\ast$ and SM backgrounds at 7 TeV and 14 TeV LHC are collected in Tables~\ref{result7} and \ref{result14} respectively.

\begin{table}[tb]
\begin{center}
\begin{tabular}[t]{|c|c|c|c|c|c|}
  \hline
  $\sigma({\rm pb})$ @ 7 TeV & $\tilde{b}_1\tilde{b}_1^\ast$ & $b\bar{b}$ & $b\bar{b}Z(\to \nu\bar{\nu})$ & $jjZ(\to \nu\bar{\nu})$ & $t\bar{t}$\\
  \hline
  basic cuts+2 b tagging & 0.22 & 7716 & 0.108 & 0.049 & 0.05\\
  \hline
  $\cancel{E}_T>40$ GeV & 0.15 & 8 & 0.07 & 0.033 & 0.029\\
  \hline
  $S_T>0.2,\Delta\phi(\cancel{E}_T,b)>0.2$ & 0.078 & - & 0.03 & 0.007 & 0.018\\
  \hline 
\end{tabular}
\end{center}
\caption{Total cross sections of $\tilde{b}_1\tilde{b}_1^\ast$ signal and SM backgrounds after different cuts at 7 TeV LHC. The masses of $\tilde{b}_1$ and $\tilde{\chi}_1^0$ are assumed to be 200 GeV and 150 GeV respectively.}
\label{result7}
\end{table}

\begin{table}[tb]
\begin{center}
\begin{tabular}[t]{|c|c|c|c|c|c|c|}
  \hline
  $\sigma({\rm pb})$ @ 14 TeV & $\tilde{b}_1\tilde{b}_1^\ast$ & $b\bar{b}$ & $b\bar{b}Z(\to \nu\bar{\nu})$ & $jjZ(\to \nu\bar{\nu})$ & $t\bar{t}$\\
  \hline
  basic cuts+2 b tagging & 0.942 & 20151 & 0.276 & 0.13 & 0.206\\
  \hline
  $\cancel{E}_T>50$ GeV & 0.51 & 12 & 0.137 & 0.052 & 0.112\\
  \hline
  $S_T>0.2,\Delta\phi(\cancel{E}_T,b)>0.2$ & 0.279 & - & 0.069 & 0.026 & 0.059\\
  \hline 
\end{tabular}
\end{center}
\caption{Total cross sections of $\tilde{b}_1\tilde{b}_1^\ast$ signal and SM backgrounds after different cuts at 14 TeV LHC. The masses of $\tilde{b}_1$ and $\tilde{\chi}_1^0$ are assumed to be 200 GeV and 150 GeV respectively.}
\label{result14}
\end{table}

\section{Results for NLSP Sbottom Discovery Limit}

\begin{figure}[tb]
\begin{center}
\includegraphics[scale=1,width=10cm]{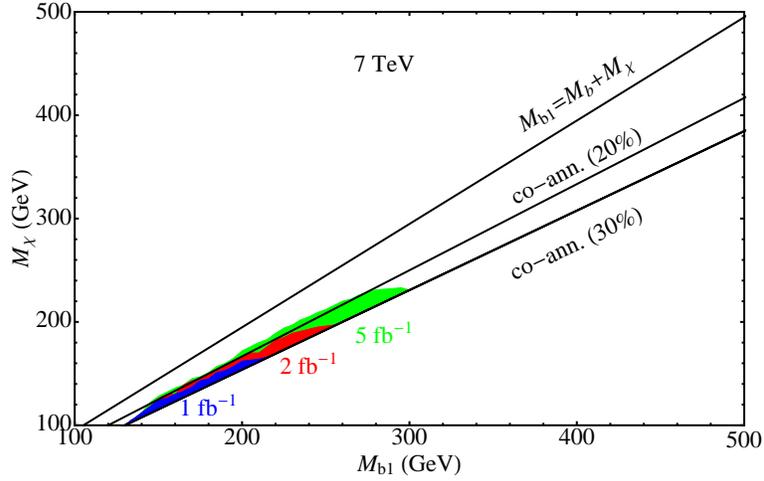}
\end{center}
\caption{Discovery region of the NLSP sbottom with $>5\sigma$ significance at 7 TeV LHC in the $M_{\tilde{b}_1}-M_{\tilde{\chi}_1^0}$ plane for integrated luminosities of 1 fb$^{-1}$, 2 fb$^{-1}$ and 5 fb$^{-1}$. The curves for the kinematic limit of $\tilde{b}_1\to b+\tilde{\chi}_1^0$ decay channel and NLSP sbottom co-annihilation requirement are also displayed.} \label{msb-mx-7}
\end{figure}

\begin{figure}[tb]
\begin{center}
\includegraphics[scale=1,width=10cm]{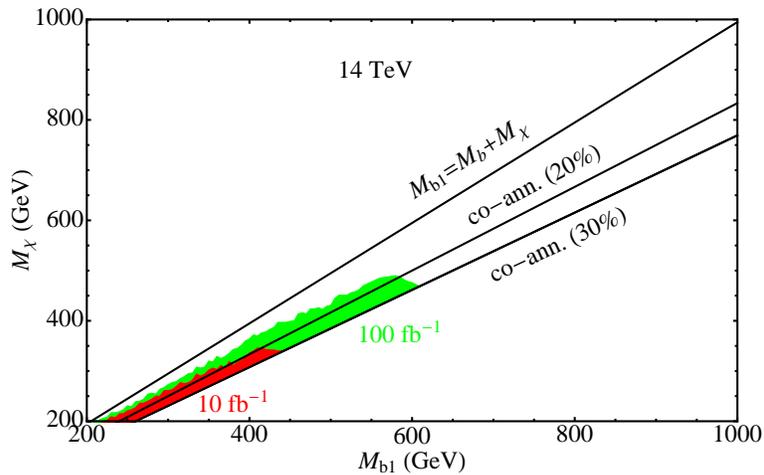}
\end{center}
\caption{Discovery region of the NLSP sbottom with $>5\sigma$ significance at 14 TeV LHC in the $M_{\tilde{b}_1}-M_{\tilde{\chi}_1^0}$ plane for integrated luminosities of 10 fb$^{-1}$ and 100 fb$^{-1}$. The curves for the kinematic limit of $\tilde{b}_1\to b+\tilde{\chi}_1^0$ decay channel and NLSP sbottom co-annihilation requirement are also displayed.} \label{msb-mx-14}
\end{figure}

As mentioned earlier, we are interested in the mass parameter region constrained by Eq.~(\ref{bound}). Thus, we scan the relevant mass parameters of $\tilde{b}_1$ and $\tilde{\chi}_1^0$ in that region. The signal significance is obtained in terms of Gaussian statistics, given by the ratio $S/\sqrt{B}$ of signal and background events with different integrated luminosities. In Figs.~\ref{msb-mx-7} and \ref{msb-mx-14} we show the discovery region of the NLSP sbottom with $>5\sigma$ significance in the $M_{\tilde{b}_1}-M_{\tilde{\chi}_1^0}$ plane at 7 TeV LHC for integrated luminosities 1 fb$^{-1}$, 2 fb$^{-1}$ and 5 fb$^{-1}$, and 14 TeV LHC for integrated luminosities 10 fb$^{-1}$ and 100 fb$^{-1}$, respectively. The straight lines correspond to the kinematic limit relation from the decay $\tilde{b}_1\to b\tilde{\chi}_1^0$, and the sbottom-bino co-annihilation requirement $\Delta M<20\%(30\%) M_{\tilde{\chi}_1^0}$. The middle region between the straight lines is what we scanned. Generally, when the parameter space extends to the kinematic limit,  the discovery sensitivity vanishes because the $b$-jets become too soft and thus cannot survive the $p_T^j$ and $\cancel{E}_T$ cuts. One can see that only with at least 5 fb$^{-1}$ luminosity the 7 TeV LHC can explore the mass region $190 \ {\rm GeV}\lesssim M_{\tilde{b}_1}\lesssim 265 \ {\rm GeV}$ satisfying the $20\%$ co-annihilation boundary, and a very narrow range above the $20\%$ co-annihilation boundary line (with width $\lesssim 5$ GeV). If we loosen the co-annihilation requirement to $30\%$, the 7 TeV LHC with luminosities 1 fb$^{-1}$, 2 fb$^{-1}$ and 5 fb$^{-1}$ can probe NLSP sbottom masses of 215 GeV, 260 GeV, and 300 GeV respectively. For 14 TeV LHC the discovery region corresponding to $20\%$ co-annihilation boundary is much broader. The relevant discovery limit is $M_{\tilde{b}_1}\sim 360$ GeV (570 GeV) for 10 fb$^{-1}$ (100 fb$^{-1}$) luminosity.

Finally, let us note that the heavy gluino pair production channel may provide an interesting avenue to search for NLSP sbottom at the LHC. As an example, consider a low-energy spectrum which has, say, $M_{\tilde{g}}\sim 1.2$ TeV and $M_{\tilde{b}_1}\sim 200$ GeV. The total production cross section for $pp\to \tilde{g}\tilde{g}X\to bb\tilde{b}_1^\ast \tilde{b}_1^\ast +\cdot\cdot\cdot$
is about 7 fb at the 14 TeV LHC, with approximately 27\% branching ratio of gluino decay into $b+\tilde{b}_1^\ast$. 
The $b$-jet and NLSP sbottom will be highly boosted, and the signal from heavy gluino decays will consequently be two extremely hard jets with one of them being the collimated sbottom. However, if we require the two energetic $b$-jets to be identified, the production rate will be further suppressed because the $b$-jet tagging efficiency declines as its $p_T$ increases, with $p_T\sim M_{\tilde{g}}/2\sim 600$ GeV in this case. It is also difficult to tag the $b$-jet in the decay products of the boosted sbottom. Therefore, the NLSP sbottom search from gluino production will require a more careful analysis of the $b$-jet structure and we leave it for future study.

\section{Conclusion}
We have explored the collider phenomenology of sbottom-bino co-annihilation scenario at the LHC. The NLSP sbottom is assumed to be slightly more massive than the LSP bino, namely $M_{\tilde{b}_1}-M_{\tilde{\chi}_1^0}\lesssim 20\% M_{\tilde{\chi}_1^0}$, with $M_{\tilde{b}_1}>M_b+M_{\tilde{\chi}_1^0}$, such that $\tilde{b}_1$ decays into $b+\tilde{\chi}_1^0$ with 100\% branching fraction. We present a search for sbottom pairs through $b\bar{b}$ plus missing energy final states. By scanning the mass parameters $M_{\tilde{b}_1}$ and $M_{\tilde{\chi}_1^0}$, we investigate the discovery limits of sbottom and bino at the 7 TeV and 14 TeV LHC with different integrated luminosities. It is shown that the 7 TeV LHC can probe only a narrow region satisfying the $20\%$ co-annihilation boundary with at least 5 fb$^{-1}$ luminosity. For 14 TeV LHC the discovery limit is $M_{\tilde{b}_1}\sim 360$ GeV and $M_{\tilde{b}_1}\sim 570$ GeV for luminosity of 10 fb$^{-1}$ and 100 fb$^{-1}$ respectively. We also briefly explored the possibility of detecting NLSP sbottom from heavy gluino decay.
\subsection*{Acknowledgment}
We would like to thank Alexander Belyaev, Ilia Gogoladze, Tao Han, Stephano Profumo and especially Kai Wang for useful discussions. This work is supported by the DOE under grant No. DE-FG02-91ER40626.



\end{document}